\documentclass[conference]{IEEEtran}
\IEEEoverridecommandlockouts

\usepackage{cite}
\usepackage{amsmath,amssymb,amsfonts}
\usepackage{algorithmic}
\usepackage{graphicx}
\usepackage{textcomp}
\usepackage{xcolor}

\usepackage{cite}
\usepackage{float}
\usepackage{pifont}
\usepackage{footnote}
\usepackage{enumitem}
\usepackage{bm}
\usepackage{arydshln}
\usepackage{booktabs}
\usepackage{multicol}
\usepackage{multirow}
\usepackage{color}
\usepackage{xcolor}     
\usepackage{colortbl}
\usepackage{soul}
\usepackage{bbding}
\usepackage{makecell}
\usepackage{mathtools}
\usepackage{imakeidx}
\usepackage{arydshln}
\usepackage{lipsum}
\usepackage[toc]{multitoc}
\usepackage[edges]{forest}
\usepackage[normalem]{ulem}

\usepackage{bbding}
\usepackage[most]{tcolorbox}

\usepackage{algorithm}
\usepackage{algorithmic}

\newcommand{\ourmodel}{{\fontfamily{ppl}\selectfont LoASP}}
\newcommand{\ouradaptation}{{\fontfamily{ppl}\selectfont LoSP}}
\newcommand{\ourfusion}{{\fontfamily{ppl}\selectfont LoAP}}
\newcommand{\etal}{\textit{et al.}}

\usepackage{pifont}       
\usepackage{adjustbox}
\usepackage{makecell}
\usepackage{amsmath}
\usepackage{bm}
\usepackage{multirow}
\usepackage{scalerel}  

\usepackage{graphicx}   

\usepackage[T1]{fontenc} 
\usepackage{amsmath}
\usepackage{bm} 

\definecolor{cvprblue}{rgb}{0.21,0.49,0.74}
\definecolor{Mahogany}{RGB}{192,64,0}
\definecolor{CadetBlue}{rgb}{0.3725, 0.6196, 0.6275}

\usepackage[pagebackref]{hyperref}
\hypersetup{
	final,
	breaklinks,
	colorlinks,
	linkcolor={Mahogany},
	citecolor={cvprblue},
	urlcolor={CadetBlue}
}

\def\BibTeX{{\rm B\kern-.05em{\sc i\kern-.025em b}\kern-.08em
    T\kern-.1667em\lower.7ex\hbox{E}\kern-.125emX}}
\begin{document}

\title{
Low-Rank Adaptive Structural Priors  \\  for  Generalizable Diabetic Retinopathy Grading 
}
\author{
\IEEEauthorblockN{
Yunxuan Wang\IEEEauthorrefmark{1},
Ray Yin\IEEEauthorrefmark{2},
Yumei Tan\IEEEauthorrefmark{3}, 
Hao Chen\IEEEauthorrefmark{4},
Haiying Xia\IEEEauthorrefmark{1}
}
\IEEEauthorblockA{\IEEEauthorrefmark{1}School of Electronic and Information Engineering, Guangxi Normal University, GuiLin, China}
\IEEEauthorblockA{\IEEEauthorrefmark{2}Nanjing First Hospital,Nanjing Medical University, NanJing, China }
\IEEEauthorblockA{\IEEEauthorrefmark{3}School of Computer Science and Engineering, Guangxi Normal University, GuiLin, China}
\IEEEauthorblockA{\IEEEauthorrefmark{4}University of Cambridge, Cambridge, UK}
\IEEEauthorblockA{Corresponding Author: Haiying Xia\IEEEauthorrefmark{1} \quad Email: xhy22@mailbox.gxnu.edu.cn}
}





\maketitle

\begin{abstract}


Diabetic retinopathy (DR), a serious ocular complication of diabetes, is one of the primary causes of vision loss among retinal vascular diseases.  {Deep learning methods have been extensively applied in the grading of diabetic retinopathy (DR). However, their performance declines significantly when applied to data outside the training distribution due to domain shifts. Domain generalization (DG) has emerged as a solution to this challenge. However, most existing DG methods overlook lesion-specific features, resulting in insufficient accuracy.} 
In this paper, we propose a novel approach that enhances existing DG methods by incorporating \ul{structural priors}, inspired by the observation that DR grading is heavily dependent on vessel and lesion structures. We introduce \textit{Low-rank Adaptive Structural Priors (LoASP)}, a plug-and-play framework designed for seamless integration with existing DG models. LoASP improves generalization by learning adaptive structural representations that are finely tuned to the complexities of DR diagnosis. Extensive experiments on eight diverse datasets validate its effectiveness in both single-source and multisource domain scenarios. Furthermore, visualizations reveal that the learned structural priors intuitively align with the intricate architecture of the vessels and lesions, providing compelling insights into their interpretability and diagnostic relevance.

%

\end{abstract}


\section{Introduction}

\noindent \textbf{Background.} By 2040, it is estimated that around 600 million people worldwide will have diabetes, and one-third of these individuals will suffer from diabetic retinopathy (DR)~\cite{dai2021DeepLearningSystem}. DR, a severe complication of diabetes, progressively damages the retina's delicate vascular networks. This damage can lead to changes such as angiogenesis and capillary regression, which may result in irreversible blindness if not timely detected and treated. Often, early-stage DR remains asymptomatic or results in minor vision disturbances, making early detection and accurate grading of DR crucial to preventing blindness~\cite{wykoff2021risk}.

Recent advancements in deep learning (DL) have significantly improved automated diagnosis in ophthalmology through medical image analysis. Current research on DR grading mainly focuses on the accuracy of detection of small lesions in the same dataset. These methods have achieved promising results in automated DR grading~\cite{li2019diagnostic,liu2020green}, but neglect the data heterogeneity between datasets, \textit{i.e.}, \textit{\textbf{data shifts}}~\cite{atwany2022drgen,che2022learning}, which results in the inability of the existing models to be directly applied to real-world clinical settings and poses a substantial obstacle in clinical practice~\cite{hu2024assessing}. Therefore, how to make models well adapted to out-of-distribution data is very crucial in clinical applications.

\begin{figure*}[t]
	\centering
\includegraphics[width=1\linewidth]{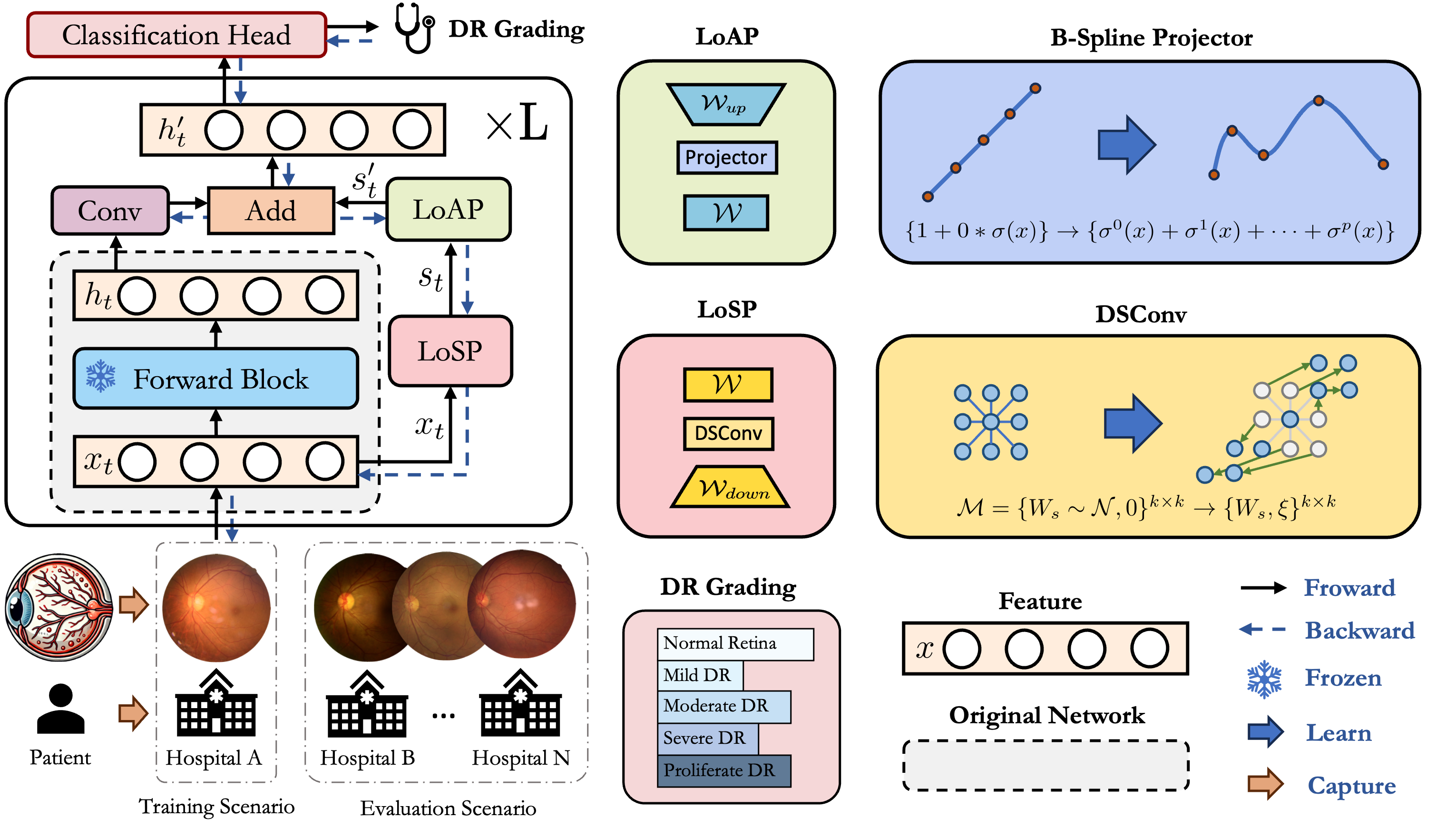}
	\caption{
    \textbf{Overview of the \ourmodel~framework.} The \ourmodel~framework comprises two key modules: the structural prior learning module (\ouradaptation) and the adaptive projection module (\ourfusion) for effective prior integration. Our method is designed to excel in limited-data training scenarios from a few source domains (hospitals), aiming for superior performance and showcasing adaptability and robustness across diverse, unseen target domains. In the main figure,  the black arrows represent the forward flow, while the blue dashed arrows illustrate the gradient backpropagation flow.
}
	\label{fig:method}
\end{figure*}


{The challenge of domain shifts has spurred the development of \textit{domain generalization} (DG), a framework designed to enhance model robustness by enabling effective generalization across diverse domains without requiring target domain data during training \cite{yang2022tally,zhang2020generalizing,zhou2022domain,chen2024domain}. In recent years, domain-invariant representations have gained attention, which employ regularization techniques to stabilize model behavior across domains \cite{atwany2022drgen, yao2023improving,chen2024domain}.  
Additionally, some studies have explored disentangling domain-specific~\cite{chokuwa2023generalizing}, domain-invariant~\cite{atwany2022drgen,che2022learning}, or intra-class features~\cite{che2023towards}. Simultaneously, image augmentation techniques have been explored to simulate domain variability, using visual transformations or deliberate image degradation to enable models to tackle the diverse challenges posed by shifting domain characteristics \cite{che2023towards,yao2022improving,yao2022c}. }




\smallskip
\noindent\textbf{Motivation.} Existing DG methods largely focus on addressing domain shifts by emphasizing data characteristics, such as pixel intensity distributions (style shifts) or data imbalance. However, we argue that these methods neglect a crucial aspect: the diagnostic features that clinicians rely on in real-world settings to grade diabetic retinopathy (DR). These features, inherently invariant across domains, are central to accurate diagnosis and serve as a robust foundation for clinical decision-making. In this work, we go beyond conventional DG strategies by focusing on these domain-invariant diagnostic features specific to DR grading. 

The International Council of Ophthalmology (ICO) defines diabetic retinopathy (DR) stages through a five-grade system (Grade 0 to 4) based on fundus diagnosis, with a particular emphasis on blood vessel-related characteristics \cite{sebastian2023survey}:
\begin{itemize}[itemsep=0.2em]
    \item {Grade 0 (Normal Retina):} \ul{\textit{Clear blood vessels}} with no visible lesions in the fundus images.
    
    \item {Grade 1 (Mild DR):} 
      Characterized by   \ul{\textit{microaneurysms}}, appearing as small dark red dots in the retina.
    
    \item {Grade 2 (Moderate DR):} 
     Presence of   \ul{\textit{microaneurysms, hemorrhages, and hard exudates}}.
    
    \item {Grade 3 (Severe DR):}
    Includes  \ul{\textit{abnormal blood vessels, extensive hemorrhages or blood clots, and soft exudates}}.
    
    \item {Grade 4 (Proliferative DR):}
 Incorporates all prior lesions along with \ul{\textit{neovascularization}}, indicating advanced disease progression.
\end{itemize}


{Building upon this, we aim to utilize vessel-related characteristics as domain-invariant diagnostic features to address the domain shift problem. In this work, we propose Low-rank Adaptive Structural Priors (LoASP), comprising three primary components: the Low-rank Structural Prior (LoSP) module, the Low-rank Adaptive Projection (LoAP) module, and the Low-rank Design Mechanism. First, LoSP dynamically captures the structural information of the vessel. Then, LoAP efficiently fuses the imaging features with the structural prior via an adaptive B-spline projector. Finally, to further cope with the challenging nature of medical datasets, we incorporate a low-rank design into our module. Notably, our proposed LoASP is not only a promising solution for domain generalization (DG) but also a plug-and-play tool that enhances the performance of existing DG methods.}

In this work, we introduce a domain-invariant structural prior to develop a novel domain generalization module, \ourmodel, which consists of two key components: \ouradaptation~and \ourfusion. This innovative framework not only functions as an efficient standalone domain generalization module but also serves as a versatile, plug-and-play enhancement that can be seamlessly integrated with existing domain generalization techniques. Its adaptability makes it a promising solution for advancing diabetic retinopathy grading through adaptive domain generalization. Specifically:

\begin{enumerate}
    \item We develop a vessel-related structural prior, termed \ouradaptation, inspired by clinical observations in DR grading. This prior inherently captures domain-invariant characteristics, enabling robust and reliable analysis (Sec.~\ref{sec:ouradaptation}).

    \item To ensure efficient prior integration with current methods, we propose \ourfusion, a lightweight adapter featuring a learnable fusion projector. This projector adaptively projects the structural prior before fusing it with the original model output, enabling a more coherent and effective synergy (Sec.~\ref{sec:ourfusion}).

    \item We verify our modules in several ways under domain generalization settings, first is the verify the module solely with model backbone, second is the to with the state-of-the-art domain generalization methods. We further done the detailed ablation into the design details of in-submodules and the parameters. We visualize the structural information and further solify our method in Sec.~\ref{sec:visualize}.
\end{enumerate}

\section{related work}

\subsection{Diabetic Retinopathy Analysis}
In recent years, deep learning has achieved remarkable advancements in DR grading, leveraging innovative architectures and methodologies to enhance diagnostic performance. Wang \etal~\cite{wang2017zoom} introduced Zoom-in-Net, an approach inspired by the clinical workflow, which utilizes attention maps to focus on suspicious regions in retinal images, mimicking how doctors examine these images. Similarly, Huang \etal~\cite{huang2021lesion} developed a self-supervised framework centered on lesion contrast learning, enabling automated DR grading by emphasizing lesion-specific features.  Expanding on these ideas, Sun \etal~\cite{sun2021lesion} proposed a transformer-based model that integrates pixel relationship-based encoders with lesion filter-based decoders, offering a unified approach to DR grading. Luo \etal~\cite{luo2020automatic} addressed the trade-off between model performance and computational efficiency through self-knowledge extraction techniques. Recognizing the clinical overlap between DR and diabetic macular edema (DME), Li \etal~\cite{li2019canet} introduced the Cross-Disease Attention Network (CANet), a model designed for the joint classification of lesions associated with both conditions.

Recently, an attention-based network called HA-Net~\cite{shaik2022hinge} was proposed to incorporate multiple attention stages to learn details in fundus images for accurate DR severity grading. In addition, the pure attention-based network, the vision transformer (ViT)~\cite{vit}, also inspired the model designing of the DR grading system. Wu \etal~\cite{vitDR} adopted ViT on public fundus datasets to achieve acceptable performance. Inspired by this, Sun \etal~~\cite{cvprsun2021lesion} proposed a lesion-aware Transformer (LAT) network for DR grading and lesion discovery.

\subsection{Domain Generalization in Medical Imaging}

Deep  models hold great promise for DR grading but often struggle with variations in imaging devices, conditions, and patient demographics across medical institutions. Domain generalization techniques address this challenge by improving model robustness and adaptability, making DR grading tools more reliable and applicable in diverse clinical settings without access to target domain. For instance, research by Zhang \etal~\cite{zhang2018mixup, zhou2020domain} introduced Mixup-based methods that linearly interpolate image-label pairs between domains, reducing spurious correlations between domain characteristics and image appearance. Nam \etal~\cite{nam2021reducing} tackled style bias by disentangling image style and content, leveraging the assumption that styles are more susceptible to domain shifts. To further learn domain-invariant representations, Sun \etal~\cite{sun2016deep} proposed minimizing feature covariance discrepancies across domains in the training set. Similarly, Li \etal~\cite{li2020DGnips} enhanced generalization by constraining the distribution of encoded features to follow a predefined Gaussian distribution.

Recent research on gradient alignment has introduced innovative methods to address inter-domain variability and improve model robustness. For example, Shi \etal~\cite{shi2021gradient} and Rame \etal~\cite{rame2022fishr} incorporated gradient matching and covariance alignment during training, promoting the learning of invariant features and reducing domain-specific biases. In parallel, optimization techniques have been employed to enhance generalization. Atwany \etal~\cite{atwany2022drgen} adopted a stochastic weight averaging schema during training, improving model performance on previously unseen datasets. Similarly, GDRNet~\cite{che2022learning} introduced a unified framework that integrates pixel-level consistency with image-level semantics to capture diverse diagnostic patterns and construct universal feature representations. Furthermore, DECO~\cite{xia2024generalizing} proposed disentangling semantic features from domain-specific noise using a pixel-level semantic registration loss, improving feature consistency across domains.

Several studies have also focused on learning causal relationships between causes and effects. For instance, Ouyang \etal~\cite{ouyang2022causality} integrated causality into a data augmentation strategy for organ segmentation, synthesizing domain-shifted training examples to improve generalization. Similarly, Chen \etal~\cite{chen2024domain} introduced a disentanglement framework that addresses the transformation sensitivity of diagnostic and domain-specific features. By leveraging causal principles, they proposed a schema with two branches: one for learning diagnostic features and another for learning domain features, thereby enhancing robustness and improving model performance across diverse domains.

\subsection{Low-rank Learning}
In this paper, we explore two widely used low-rank learning frameworks: Low-Rank Adaptation (LoRA)~\cite{hu2021lora} and Adapter~\cite{houlsby2019parameter}. LoRA, originally developed for efficient fine-tuning in natural language processing, has rapidly gained popularity in vision tasks. Its core innovation lies in enabling network adaptation without directly modifying the pre-trained weight matrix $ \mathcal W_0 $. Specifically, LoRA introduces a low-rank learnable module that operates alongside $\mathcal W_0 $, leveraging the observation that high-dimensional weight matrices can be effectively approximated in a low-dimensional space. During fine-tuning, LoRA represents the weight update $ \Delta \mathcal W $ as the product of two low-rank matrices $ A $ and $ B $, such that the updated weight becomes:

\begin{equation}
\mathcal W' = \mathcal W_0 + BA
\end{equation}

Here, $ A $ and $ B $ are trainable matrices, and their dimensions are significantly smaller than those of $ \mathcal W_0 $, reflecting the low-rank approximation. The resulting forward pass can be expressed as:

\begin{equation}
h_{t}' = (\mathcal W_0 + BA)\ast x_t
\end{equation}

where $ x_t $ and $h_t'$ are the input and output at layer block $ t $, with the superscript $'$ indicating the modified version of original output $h_t$. The operation $\ast$ indicates matrix multiplication.  In this formulation, matrix $ A $ is initialized with Gaussian random values, while $ B$ is zero-initialized. This setup ensures that at the initial step, the output remains consistent with the pre-trained model, $ h_{t}' = h_{t} = \mathcal W_0\ast x_t $, preserving the original model's behavior and maintaining the initial layer output $h_{t+1}$.

Similarly, the Adapter fine-tuning technique reduces the number of trainable parameters during fine-tuning, while maintaining model performance. This is achieved by inserting lightweight modules sequentially into the pre-trained model, as opposed to the parallel approach employed by LoRA. Adapter modules typically consist of an input layer, an output layer, a lower projection feedforward layer, an upper projection feedforward layer, a nonlinear activation layer, and a skip connection from input to output.

The operational principle of the Adapter can be expressed as:

\begin{equation}
h_{t}' = h_t + B \ast\sigma(A\ast h_t), 
\end{equation}

\noindent where $ \sigma $ represents the nonlinear activation function,  $ A $ and $ B $ are down- and up-scale projection matrices. To ensure that the Adapter module behaves as an identity mapping at the beginning of fine-tuning, the matrices $ A $ and $ B $ are initialized with near-zero values.

\section{Methodology}
\label{sec:method}

According to observation, vascular structure plays a critical role in DR Classification, particularly for small lesions. However, existing domain adaptation methods for DR neglect small lesions. Thus, we propose LoASP, which incorporates domain-invariant structural priors to learn robust representations to reduce the discrepancy between the source and target domains.  As shown in  \figureautorefname~\ref{fig:method}, \ourmodel~ contains two parts: \ouradaptation~ and \ourfusion~. \ouradaptation~ for structure learning and  \ourfusion~ for structural information fusion. Above all, we use an overall low-rank design to reduce computational complexity and overfitting.



\subsection{\ouradaptation}
\label{sec:ouradaptation}
Traditional convolutional layers rely on fixed convolution kernels for feature extraction, which struggle to capture the complex tubular structures, curved microvessels, and small lesion areas in fundus images. To address this issue, we propose the LoSP to capture complex structural information by adaptively focusing on local features through dynamic convolution kernels. 

Specifically, LoSP introduces Dynamic Snake Convolution  (DSConv) \cite{qi2023dynamic} to learn structural priors and adaptively focus on the small, curved local features of tubular structures, thereby improving the model's ability to capture complex structural features. \textit{DSConv} introduces learnable position Cartesian offsets ($ \xi $), enabling the kernel shape to adapt dynamically during training. This enhancement allows DSConv to better capture spatial features and structural variations. The kernel learning process can be expressed as:

\begin{equation}
  \mathcal M =  \{W_s \sim \mathcal{N}, 0\}^{k \times k} \to \{W_s, \xi\}^{k \times k}, 
\end{equation}

where $\mathcal M $ represents the kernel matrix, $ W_s $ denotes the kernel weights initialed with a standrad Gaussian distribution $\mathcal{N}$, the offset is initialized to 0 and $ k \times k $ indicates the kernel size. By learning offsets dynamically, DSConv enhances the model's capacity to adapt to diverse spatial patterns, improving its ability to extract meaningful structure.

However, directly employing DSConv can involve two risks that result in performance drops as demonstrated in Sec.~\ref{sec:ablation}: overfitting and optimization challenges. The added flexibility of dynamic kernels may lead to overfitting, particularly when applied to small or less diverse medical datasets. Additionally, the increased offset parameterization of dynamic kernels can make optimization more challenging. Thus, \ouradaptation incorporates low-rank learning strategy to address the aforementioned problems. Finally, the \ouradaptation~can be expressed as:

\begin{equation}
s_{t} =  B_s \ast  f(A_s\ast x_t, \mathcal W_s, \xi),
\end{equation}

where \( s_t \) represents the learned structural prior. \( A_s \) serves as the down-scaling projection weight, reducing dimensionality, while \( B_s \) maintains the same resolution, acting as a same-scale projection matrix. Each projector are embedded within a convolutional layer, followed by batch normalization for stable learning.

\subsection{\ourfusion}
\label{sec:ourfusion}

To better fuse structural information, we propose the \textit{Low-Rank Adaptive Projector }(\ourfusion). \ourfusion~incorporates a dynamic non-linearity module \(\sigma(\cdot, u)\), computed in a resolution-wise  manner. This module leverages the B-Spline function~\cite{schoenberg1988contributions}, where \(u\) defines the parameters of the non-linear mapping. The output is formally expressed as:
\begin{equation}
s_{t}' = B_f \ast  \sigma^p(A_f\ast  s_{t}, u),
\end{equation}
\noindent where \( A_f \) is a same-scale projection weight, and \( B_f \) serves as the up-scaling projection weight. Then, the block output $h_t'$ is computed as:
\begin{equation}
h_{t}' = (A_c \ast h_{t}) + h_{t}  + s_{t}',
\end{equation}
where \( A_c \) represents the kernel matrix of an additional convolutional layer designed to refine the original \( h_t \).

The B-spline degree and the control point parameters are denoted as \( p \) and \( u \), respectively.  The construction of basis functions begins with the zeroth-degree (\( p\!=\!0 \)) case, corresponding to a piecewise constant function. The basis function \(\sigma_i^p(x)\) is defined as:

\begin{equation}
\sigma_i^p(x) =
\begin{cases} 
1, & \text{if } u_i \leq x < u_{i+1}, \\
0, & \text{otherwise},
\end{cases}
\quad \text{for } p = 0.
\end{equation}

 For higher-degree functions (\(p > 0\)), the construction relies on a recurrence relation. This relation elegantly combines two weighted components, with the weights determined by the relative position of \(x\) within the intervals \([u_i, u_{i+p}]\) and \([u_{i+1}, u_{i+p+1}]\). The recursive definition is given by:

\begin{equation}
\sigma_i^p(x) = \frac{x - u_i}{u_{i+p} - u_i} \sigma_i^{p-1}(x) + \frac{u_{i+p+1} - x}{u_{i+p+1} - u_{i+1}} \sigma_{i+1}^{p-1}(x).
\end{equation}

\begin{table*}[t]
\centering
\footnotesize
\caption{Performance comparison of \ourmodel~against state-of-the-art methods in the domain generalization setting, integrated into GDRNet~\cite{che2023towards}, Domain Game~\cite{chen2024domain}, and DECO~\cite{xia2024generalizing}, with the suffix (\ourmodel) indicating its inclusion.}
\label{tab:dg}
\resizebox{\linewidth}{!}{
\begin{tabular}{l|ccc|ccc|ccc|ccc|ccc|ccc|ccc}
\toprule
Target & \multicolumn{3}{c|}{APTOS~\cite{karthick2019aptos}} & \multicolumn{3}{c|}{DeepDR~\cite{liu2022deepdrid}} & \multicolumn{3}{c|}{FGADR~\cite{zhou2020benchmark}} & \multicolumn{3}{c|}{IDRID~\cite{porwal2018indian}} & \multicolumn{3}{c|}{Messidor~\cite{abramoff2016improved}} & \multicolumn{3}{c|}{RLDR~\cite{wei2021learn}} & \multicolumn{3}{c}{Average} \\ 

\midrule Metrics & AUC  & ACC  & F1  & AUC & ACC & F1 & AUC & ACC & F1 & AUC & ACC & F1 & AUC & ACC & F1 & AUC & ACC & F1 & AUC & ACC & F1 \\ \midrule 
ERM & 77.02 & 46.64 & 39.43 & 76.31 & 39.15 & 35.14 & 66.08 & 30.63 & 26.39 & 83.96 & \textbf{52.17} & \textbf{42.10} & 77.76 & 59.54 & 41.35 & 74.38 & 35.24 & 35.72 & 75.92 & 43.89 & 36.69 \\
Mixup~\cite{zhang2018mixup} & 75.90 & 63.62 & 45.10 & 75.97 & 27.31 & 26.90 & 65.66 & 44.38 & 31.94 & 81.81 & 36.68 & 28.71 & 76.23 & 56.22 & 33.45 & 79.15 & 43.98 & 36.57 & 75.79 & 45.37 & 33.78 \\
MixStyle~\cite{zhou2020domain} & 79.17 & 64.86 & 37.36 & 78.95 & 34.29 & 25.91 & 71.58 & 35.36 & 24.10 & 85.53 & \ul{51.02} & \ul{41.56} & 76.30 & 63.29 & 37.87 & 75.38 & 39.93 & 30.73 & 77.82 & 48.12 & 32.92 \\
GREEN~\cite{liu2020green} & 75.59 & 51.36 & 38.10 & 74.96 & 26.50 & 24.35 & 70.24 & 41.09 & 31.34 & 78.39 & 39.74 & 32.74 & 77.53 & 54.80 & 37.16 & 73.31 & 33.16 & 35.51 & 75.00 & 41.11 & 33.20 \\
CABNet~\cite{he2020cabnet} & 75.72 & 53.92 & 39.18 & 76.03 & 40.88 & 31.71 & 73.69 & 42.52 & 32.24 & 79.66 & 44.71 & 36.17 & 71.94 & 54.77 & 32.57 & 74.14 & 38.80 & 36.77 & 75.20 & 45.93 & 34.77 \\
DDAIG~\cite{zhou2020deep} & 79.00 & 69.62 & 41.72 & 76.74 & 36.27 & 31.33 & 72.62 & 40.10 & 35.10 & 84.99 & 36.55 & 27.40 & 74.15 & 59.62 & 35.68 & 74.67 & 34.64 & 28.10 & 77.03 & 46.13 & 33.22 \\
ATS~\cite{yang2021adversarial} & 76.28 & 58.00 & 35.07 & 78.34 & 36.07 & 31.52 & 74.58 & 47.24 & 32.72 & 80.54 & 42.50 & 35.11 & 75.62 & 64.28 & 37.90 & 76.33 & 37.22 & 34.72 & 76.95 & 47.55 & 34.51 \\
Fishr~\cite{rame2022fishr} & 81.54 & 64.53 & 42.00 & 81.77 & 49.95 & 34.51 & 71.24 & 45.41 & 37.15 & 85.44 & 40.12 & 28.32 & 78.88 & 65.69 & 41.99 & 77.30 & 36.04 & 34.37 & 79.36 & 50.29 & 36.39 \\
MDLT~\cite{yang2022multi} & 75.64 & 57.21 & 41.36 & 80.35 & 37.57 & 33.50 & 74.85 & 46.93 & 27.36 & 82.09 & 44.85 & 37.89 & 73.82 & 57.30 & 39.90 & 75.62 & 34.92 & 36.01 & 77.06 & 46.46 & 36.00 \\
DRGen~\cite{atwany2022drgen} & 81.92 & 58.43 & 39.53 & 85.12 & 40.31 & 32.84 & 70.23 & 42.26 & 27.29 & \ul{86.34} & 47.72 & 38.22 & 79.14 & 61.91 & 41.20 & 78.10 & 40.44 & 35.04 & 80.14 & 48.51 & 35.69 \\
VAE-DG~\cite{chokuwa2023generalizing} & 80.34 & 65.59 & 44.61 & 81.19 & 38.11 & 36.89 & 77.59 & 44.53 & 37.28 & 82.98 & 42.60 & 36.18 & 77.84 & 58.41 & 39.37 & 81.03 & 38.56 & 39.22 & 80.16 & 47.97 & 38.93 \\

GDRNet~\cite{che2023towards} & 78.73 & 67.30 & 47.57 & 82.99 & 54.01 & 46.58 & 82.85 & 45.13 & 40.23 & 83.00 & 38.73 & 36.81 & 82.88 & 65.67 & 49.20 & 84.20 & 45.33 & 46.02 & 82.44 & 52.69 & 44.40 \\
Domain Game~\cite{chen2024domain} & 81.32 & 69.17 & 49.02 & 82.67 & 54.46 & 47.93 & 82.41 & 47.59 & 40.82 & 84.13 & 43.66 & 37.18 & 83.21 & 66.18 & 53.79 & 84.17 & 46.96 & 46.30 & 82.98 & 54.67 & 45.84 \\
DECO~\cite{xia2024generalizing} & 81.81 & 68.08 & \ul{53.62} & 84.43 & 55.11 & 48.82 & 83.04 & \ul{49.48} & 41.33 & 85.53 & 45.36 & 39.19 & 81.05 & 66.01 & 54.05 & 83.90 & 47.14 & 46.14 & 83.29 & 55.20 & \ul{47.19} \\
\midrule
GDRNet (\ourmodel) & 81.34 & \ul{69.91} & 48.10 & 84.38 & 54.36 & \ul{49.52} & \ul{85.08} & 46.25 & \ul{42.06} & 82.98 & 39.44 & 38.89 & \ul{83.63} & 66.47 & 49.16 & \textbf{86.66} & 46.51 & 46.29 & 84.01 & 53.82 & 45.67 \\
Domain Game (\ourmodel) & \textbf{83.32} & \textbf{70.23} & 49.34 & \textbf{85.90} & \textbf{57.53} & 49.23 & 84.99 & 49.29 & 41.96 & 86.04 & 44.94 & 38.16 & \textbf{84.94} & \textbf{67.69} & \ul{54.45} & 85.36 & \ul{48.55} & \textbf{48.10} & \textbf{85.09} & \ul{56.37} & 46.87 \\
DECO (\ourmodel) & \ul{82.34} & 69.10 & \textbf{55.73} & \ul{85.52} & \ul{56.46} & \textbf{50.86} & \textbf{85.50} & \textbf{51.38} & \textbf{42.93} & \textbf{87.61} & 46.24 & 39.56 & 83.04 & \ul{66.73} & \textbf{54.67} & \ul{86.42} & \textbf{50.20} & \ul{46.81} & \ul{85.07} & \textbf{56.69} & \textbf{48.43} \\

\bottomrule
\end{tabular}}
\label{tab:dg}
\vspace{-1em}
\end{table*}

\subsection{Complexity}
We propose integrating \ourmodel~as a block-wise plug-and-play module. Let \( K,C_{\text{in}} , C_{\text{out}}, H, W\) be kernel size,  input and output channels, the spatial dimensions of the feature maps, respectively. Given rank \( r \), the complexity of projection layers is computed as:
\begin{equation}
    \mathcal{O}_{\text{proj}} = \mathcal{O}\left(K^2 \cdot C_{\text{in}} * C_{\text{out}}\cdot \frac{H}{r} \cdot \frac{W}{r}\right).
\end{equation}

For DSConv, the complexity remains similar to projection layers but includes additional operations due to extra kernel computations along both axes (\( x, y \)), contributing approximately \( 3 \times\mathcal{O}_{\text{proj}} \). Additionally, the \textit{B-Spline operation} in \ourmodel, which operates resolution-wise (\( H \times W \)), has a complexity influenced by the spline order \( p \) and the number of control points \( u \):

\begin{equation}
\mathcal{O}_{\text{spline}} = \mathcal{O}(H \cdot W \cdot (p + 1) \cdot u).
\end{equation}

Since \( (p + 1) \cdot u \) is typically much smaller than \( K^2 \cdot C_{\text{in}} \cdot C_{\text{out}} \), the spline computation imposes negligible overhead.  By combining these components, the total computational complexity of \ourmodel~can be expressed as:

\begin{equation}
\mathcal{O}_{\ourmodel} = \mathcal{O}\left(5 \cdot K^2 \cdot C_{\text{in}}  * C_{\text{out}} \cdot \frac{H}{r} \cdot \frac{W}{r}\right).
\end{equation}

Assuming \( r \geq 3 \), the computational complexity will be smaller than that of a convolutional layer, highlighting the efficiency of \ourmodel. As a benchmark, ResNet-50~\cite{he2016deep} consists of approximately 25.6M parameters. When \( r=4 \), integrating \ourmodel~into all 16 blocks of ResNet-50 adds approximately 5.9M parameters. 












\section{Experiments}

\subsection{Experimental Design}

We conduct a comprehensive evaluation of the generalization capability of our approach using two experimental protocols: the leave-one-domain-out protocol (DG)~\cite{che2023towards} and Single Domain Generalization (SGD). Following the dataset selection criteria of GDR-Net~\cite{che2023towards}, we utilize six widely used datasets for training and evaluation: DeepDR~\cite{liu2022deepdrid}, Messidor~\cite{abramoff2016improved}, IDRID~\cite{porwal2018indian}, APTOS~\cite{karthick2019aptos}, FGADR~\cite{zhou2020benchmark}, and RLDR~\cite{wei2021learn}.  For the SGD protocol, we train the model on a single dataset from the aforementioned selection and evaluate it on two large-scale datasets, DDR~\cite{li2019diagnostic} and EyePACS~\cite{emma2015eyepacs}, to assess its robustness and generalization across unseen domains. 

Additionally, we test the low-rank tuning protocol on selected methods by integrating our proposed \ourmodel~in a plug-and-play manner under the aforementioned settings, aiming to assess its compatibility and effectiveness in enhancing the performance of these approaches.

\smallskip
\noindent \textbf{Implementation Details.}
We use ResNet-50~\cite{he2016deep} as the backbone, initialized with weights pre-trained on the ImageNet dataset~\cite{deng2009imagenet}. \ourmodel~can be easily applied to \textit{Domain Game}~\cite{chen2024domain} or other domain generalization (DG) methods. Specifically, we apply \ourmodel~to the feature learning and diagnostic learning networks, \textit{e.g.,} the diagnostic branch in \textit{Domain Game}. 
For both DG and SDG settings, we train the model for 500 epochs using the AdamW optimizer, with a learning rate of $ 3 \times 10^{-3} $, a weight decay of $ 10^{-4} $, and a batch size of 32. To facilitate effective learning, we apply a step learning rate scheduler, halving the learning rate every 100 epochs.
 For the low-rank tuning protocol, the learning rate is reduced to $ 5 \times 10^{-4} $ to ensure training stability under the low-rank constraints. All hyperparameters are fine-tuned using cross-validation, combined with a grid search strategy.

\smallskip
\noindent \textbf{Metrics.} We report three essential metrics to comprehensively assess model grading performance: accuracy (ACC), the area under the ROC curve (AUC), and the macro F1-score (F1). 

\smallskip
\noindent \textbf{Remark:} The best and second-best results in the tables are highlighted in \textbf{bold} and \ul{underlined}, respectively.

\begin{table*}[t]
\centering
\footnotesize
\caption{Performance comparison against state-of-the-art methods in the challenging single-domain generalization test.}
\label{tab:esdg}
\resizebox{\linewidth}{!}{
\begin{tabular}{l|ccc|ccc|ccc|ccc|ccc|ccc|ccc}
\midrule
Source & \multicolumn{3}{c|}{APTOS~\cite{karthick2019aptos}} & \multicolumn{3}{c|}{DeepDR~\cite{liu2022deepdrid}} & \multicolumn{3}{c|}{FGADR~\cite{zhou2020benchmark}} & \multicolumn{3}{c|}{IDRID~\cite{porwal2018indian}} & \multicolumn{3}{c|}{Messidor~\cite{abramoff2016improved}} & \multicolumn{3}{c|}{RLDR~\cite{wei2021learn}} & \multicolumn{3}{c}{Average} \\ 
\midrule Metrics & AUC & ACC & F1 & AUC & ACC & F1 & AUC & ACC & F1 & AUC & ACC & F1 & AUC & ACC & F1 & AUC & ACC & F1 & AUC & ACC & F1 \\ \midrule
ERM & 64.25 & 54.15 & 32.26 & 68.72 & 45.90 & 30.22 & 55.31 & 4.00 & 6.87 & 66.91 & 55.86 & 34.78 & 68.50 & 50.17 & 31.65 & 70.09 & 28.68 & 24.67 & 65.63 & 39.79 & 26.74 \\
Mixup~\cite{zhang2018mixup} & 64.03 & 48.92 & 32.68 & 71.99 & 51.47 & 32.94 & 56.44 & 5.83 & 8.26 & 69.84 & 62.73 & 33.91 & 71.61 & 62.55 & 31.55 & 72.08 & 25.99 & 27.69 & 67.66 & 42.91 & 27.84 \\
MixStyle~\cite{zhou2020domain} & 61.60 & 49.45 & 23.22 & 53.25 & 32.61 & 13.19 & 51.94 & 7.86 & 8.85 & 50.68 & 51.95 & 18.22 & 52.37 & 58.57 & 17.61 & 53.90 & 18.55 & 6.20 & 53.96 & 36.50 & 14.55 \\
GREEN~\cite{liu2020green} & 70.43 & 53.62 & 34.82 & 74.11 & 42.54 & 33.05 & 61.08 & 3.49 & 6.40 & 68.46 & 60.75 & 32.52 & 72.69 & 55.10 & 33.54 & 73.22 & 33.29 & 28.05 & 70.00 & 41.46 & 28.06 \\
CABNet~\cite{he2020cabnet} & 68.50 & 53.34 & 28.26 & 67.30 & \ul{55.94} & 32.76 & 59.00 & 6.26 & 4.80 & 66.14 & 64.60 & 31.68 & 72.75 & 62.51 & 35.19 & 73.43 & 24.62 & 24.36 & 67.85 & 44.54 & 26.18 \\
DDAIG~\cite{zhou2020deep} & 67.78 & 48.70 & 32.76 & 73.60 & 37.37 & 30.19 & 61.22 & 4.12 & 6.05 & 69.62 & 58.99 & 33.27 & 74.28 & 70.83 & 35.33 & 74.61 & 25.02 & 23.58 & 70.19 & 40.84 & 26.86 \\
ATS~\cite{yang2021adversarial} & 67.69 & 51.92 & 32.52 & 71.25 & 54.65 & 35.46 & 62.00 & 3.36 & 3.09 & 67.31 & 68.58 & 31.04 & 73.19 & 65.83 & 34.67 & 77.43 & 24.34 & 25.22 & 69.81 & 44.78 & 27.00 \\
Fishr~\cite{rame2022fishr} & 64.77 & \ul{60.91} & 31.20 & 71.88 & \textbf{60.03} & 30.57 & 55.94 & 7.03 & 8.38 & 72.13 & 46.20 & 30.02 & 73.58 & 52.80 & 35.29 & 78.77 & 18.30 & 21.00 & 69.51 & 40.88 & 26.08 \\
MDLT~\cite{yang2022multi} & 67.00 & 54.15 & 33.72 & 75.74 & 52.55 & 34.13 & 56.39 & 8.81 & 6.80 & 72.10 & 61.36 & 32.04 & 73.90 & 58.70 & 36.91 & 76.04 & 29.25 & 29.53 & 70.20 & 44.14 & 28.86 \\
DRGen~\cite{atwany2022drgen} & 70.82 & 60.31 & 36.12 & \textbf{80.92} & 38.93 & 33.06 & 60.70 & 9.21 & 9.24 & 72.57 & 67.69 & 30.52 & 76.05 & 65.09 & 37.19 & 77.81 & 17.96 & 17.94 & 73.14 & 43.20 & 27.34 \\
VAE-DG~\cite{chokuwa2023generalizing} & 69.31 & 49.20 & 32.49 & 72.97 & 53.38 & 31.15 & 60.74 & 3.39 & 4.01 & 72.35 & 65.46 & 32.80 & 76.40 & 60.57 & 34.08 & 76.33 & 23.25 & 23.37 & 71.35 & 42.54 & 26.32 \\
GDRNet~\cite{che2023towards} & 71.11 & 51.83 & 35.60 & 75.07 & 40.77 & 34.81 & 63.66 & 7.24 & 7.90 & 73.38 & 70.82 & 35.26 & 78.38 & 65.17 & 41.16 & 78.82 & 42.86 & 37.64 & 73.40 & 46.45 & 32.06 \\
Domain Game~\cite{chen2024domain} & 72.24 & 58.17 & \ul{36.13} & 78.02 & 49.24 & \ul{38.68} & 64.29 & 9.81 & 9.27 & 75.44 & 72.09 & 36.89 & 80.83 & 70.65 & 44.12 & 80.10 & 48.37 & 38.26 & 75.15 & 51.39 & 33.89 \\
DECO~\cite{xia2024generalizing} & 71.45 & 60.86 & 34.27 & 79.20 & 39.72 & 37.41 & 64.54 & 9.34 & \ul{11.14} & 74.13 & \ul{73.54} & \ul{38.52} & 80.72 & \ul{71.88} & \ul{45.59} & 81.05 & \ul{49.92} & 37.96 & 75.18 & 50.88 & 34.15 \\
\midrule
GDRNet (\ourmodel) & 72.69 & 53.97 & 36.03 & 78.47 & 43.45 & 36.15 & 65.31 & 7.48 & 8.54 & 74.88 & 72.80 & 37.28 & 80.91 & 68.22 & 43.21 & 80.74 & 43.74 & \textbf{39.28} & 75.50 & 48.28 & 33.41 \\
Domain Game (\ourmodel) & \textbf{74.60} & 58.72 & \textbf{37.25} & 79.08 & 50.13 & \textbf{39.22} & \ul{66.04} & \ul{9.84} & 10.11 & \ul{77.49} & 73.21 & 38.37 & \textbf{83.39} & 71.23 & 45.55 & \ul{82.30} & 48.97 & 38.47 & \ul{77.15} & \ul{52.02} & \ul{34.83} \\
DECO (\ourmodel) & \ul{73.90} & \textbf{62.69} & 35.22 & \ul{79.94} & 39.54 & 37.21 & \textbf{67.80} & \textbf{10.53} & \textbf{12.12} & \textbf{77.65} & \textbf{75.88} & \textbf{40.29} & \ul{82.49} & \textbf{72.27} & \textbf{47.07} & \textbf{83.86} & \textbf{52.20} & \ul{38.55} & \textbf{77.61} & \textbf{52.18} & \textbf{35.08} \\

\bottomrule

\end{tabular}}
\label{tab:sdg}
\vspace{-1em}
\end{table*}

\subsection{Model Generalization Experiments}

We follow the experimental framework outlined in~\cite{xia2024generalizing}, which includes a diverse set of algorithms designed to benchmark performance. The Empirical Risk Minimization (ERM) approach is chosen as the traditional baseline method. To enable a nuanced evaluation of performance across multiple dimensions, we incorporate DR grading methods~\cite{he2020cabnet,liu2020green}, feature learning strategies~\cite{yang2022multi,rame2022fishr,chen2024domain,xia2024generalizing}, and domain generalization techniques~\cite{atwany2022drgen,zhang2018mixup,zhou2020deep,zhou2020domain,yang2021adversarial,chokuwa2023generalizing,che2023towards}.

\smallskip
\noindent\textbf{Domain Generalization Scenarios.} 
\tableautorefname~\ref{tab:dg} highlights the exceptional generalization performance of our approach under domain generalization (DG) settings, surpassing previous state-of-the-art (SoTA) methods across all evaluated leave-one-out datasets, based on training on the remaining datasets.
To evaluate the impact of our \ourmodel, we integrate it with three leading DG frameworks, based on their strong baseline performance: GDRNet~\cite{che2023towards}, Domain Game~\cite{chen2024domain}, and DECO~\cite{xia2024generalizing}. This decision to adopt the leading domain generalization (DG) approach stems from their potential to achieve optimal performance in real-world clinical applications. 

As shown in \tableautorefname~\ref{tab:dg}, our \ourmodel~achieves substantial average improvements of \textit{5.46} in AUC, \textit{4.32} in ACC, and \textit{3.54} in F1 score. These gains highlight the transformative potential of \ourmodel~in fruther enhancing current DG methods.

\smallskip
\noindent\textbf{Generalization in Single-Source Domain Scenarios.} 
\tableautorefname~\ref{tab:sdg} presents an in-depth analysis of generalization performance within the single domain generalization (SDG) framework. This evaluation assesses the model's ability to adapt by training exclusively on single-domain datasets and testing its performance on diverse, large-scale unseen datasets, thereby highlighting its robustness and versatility in handling real-world variability under challenging training scenarios.

Notably, the shift from the DG settings to the SDG scenarios introduces considerable challenges, as evidenced by performance declines across all evaluated methods. Despite this, GDRNet~\cite{che2023towards}, Domain Game~\cite{chen2024domain}, and DECO~\cite{xia2024generalizing} consistently demonstrate leading performance within the SDG framework. When further augmented with \ourmodel, these methods achieve remarkable improvements, with average gains of \textit{6.53} in AUC, \textit{3.76} in ACC, and \textit{3.22} in F1 score. These results highlight the robustness and efficacy of our approach in enhancing generalization under challenging conditions.


\begin{table}[t]
\centering
\footnotesize
\caption{Ablation studies on the proposed components in the DG setting, reported with ACC scores using the DECO (\ourmodel).}
\label{tab:ablation_dg}
\resizebox{\linewidth}{!}{
\begin{tabular}{c|c|cccccc|c}
\toprule
Priors & Fusion & APTOS & DeepDR & FGADR & IDRID & Messidor & RLDR & Average   \\  
\midrule
- & - & 81.81 & 84.43 & 83.04 & 85.53 & 81.05 & 83.90 & 83.29   \\
LoRA & ADD & 78.23 & 81.52 & 79.12 & 81.76 & 77.31 & 81.06 & 79.83 \\
DSConv & ADD & 63.70 & 65.23 & 61.49 & 63.40 & 61.39 & 64.97 & 63.36 \\
\ourmodel & ADD & 72.11 & 74.82 & 73.84 & 75.95 & 71.39 & 74.69 & 73.80 \\
LoRA & Adapter & 80.82 & 83.57 & 83.11 & 83.27 & 80.42 & 83.31 & 82.42 \\
DSConv & Adapter & 72.42 & 77.39 & 74.01 & 74.56 & 75.59 & 75.73 & 74.95 \\
\ourmodel & Adapter & \ul{82.17} & \ul{85.03} & 83.91 & \ul{86.29} & \ul{81.73} & \ul{84.70} & \ul{83.97} \\
LoRA & \ourfusion & 81.17 & 83.96 & \ul{84.06} & 85.37 & 81.17 & 83.96 & 83.28 \\
DSConv & \ourfusion & 79.83 & 81.30 & 79.55 & 83.04 & 79.90 & 83.62 & 81.21 \\
\midrule
\ourmodel & \ourfusion & \textbf{82.34} & \textbf{85.52} & \textbf{85.50} & \textbf{87.61} & \textbf{83.04} & \textbf{86.42} & \textbf{85.07} \\
\bottomrule
\end{tabular}
}
\label{tab:abalation}
\end{table}

\subsection{Ablation Study}
\label{sec:ablation}
\noindent
\textbf{Ablation Study of Model Components.}  
We perform an ablation study on the proposed components, \ourmodel~and \ourfusion, which serve as the prior and fusion methods, respectively. For a comprehensive comparison, we also evaluate LoRA and DSConv as prior extraction methods, and ADD operation and Adapter as alternative fusion methods. We conduct ablation experiments across both DG and SDG settings to thoroughly evaluate performance. The results, detailed in \tableautorefname~\ref{tab:ablation_dg} and \tableautorefname~\ref{tab:ablation_sdg}, showcase intriguing insights into each configuration. The use of LoRA in this context indicates the exclusion of DSConv within \ouradaptation.

As illustrated in the table, employing \textbf{LoRA} with the ADD mechanism leads to performance drops of \textit{3.46} and \textit{3.26} in the DG and SDG settings, respectively. While replacing ADD with Adapter and \ourfusion~ mitigates this decline, the use of LoRA alone remains inadequate to fully address the challenges of improving generalization capabilities. 

Moreover, the results on \textbf{DSConv} reveal a consistent pattern: using ADD for feature fusion causes substantial {performance drops} of \textit{20.05} and \textit{26.69} in DG and SDG settings. Replacing ADD with Adapter reduces these gaps to \textit{8.34} and \textit{18.51}, while \ourfusion~further improves narrow the performance drop to \textit{2.08} and \textit{8.89}. This steady progression underscores the effectiveness of \ourfusion~as a robust fusion strategy for improving generalization across diverse settings.

Another noticeable trend is that \textbf{\ourmodel}~consistently outperforms LoRA and DSConv across all fusion methods. Interestingly, combining \ourmodel~with ADD results in performance drops of \textit{4.49} and \textit{9.71} in DG and SDG settings, respectively. However, the introduction of Adapter marks a turning point, achieving gains of \textit{0.68} and \textit{0.65} over the baseline, respectively. Finally, leveraging \ourfusion~maximizes performance, delivering improvements of \textit{1.78} and \textit{2.43}, solidifying \ourmodel~as the standard configuration of our plug-and-play framework.



\begin{table}[t]
\centering
\scriptsize
\caption{Ablation studies on the proposed components in the SDG setting, reported with ACC scores using the DECO (\ourmodel).}
\label{tab:ablation_sdg}
\resizebox{\linewidth}{!}{
\begin{tabular}{c|c|cccccc|c}
\toprule
Priors & Fusion & APTOS & DeepDR & FGADR & IDRID & Messidor & RLDR & Average \\  
\midrule
- & - & 71.45 & 79.20 & 64.54 & 74.13 & 80.72 & 81.05 & 75.18   \\
LoRA & ADD & 68.46 & 75.61 & 61.16 & 71.00 & 78.43 & 76.89 & 71.92 \\
DSConv & ADD & 48.94 & 48.07 & 40.18 & 46.27 & 52.49 & 55.00 & 48.49 \\
\ourmodel & ADD & 61.77 & 69.35 & 54.91 & 64.31 & 71.04 & 71.43 & 65.47 \\
LoRA & Adapter & 69.26 & 77.23 & 63.01 & 71.50 & 80.80 & 78.40 & 73.37 \\
DSConv & Adapter & 54.30 & 58.98 & 47.08 & 60.04 & 56.28 & 63.34 & 56.67 \\
\ourmodel & Adapter & \ul{71.81} & \ul{79.90} & \ul{65.42} & 74.84 & 81.44 & \ul{81.58} & \ul{75.83} \\
LoRA & \ourfusion & 70.87 & 78.00 & 63.34 & \ul{74.96} & \ul{81.80} & 79.50 & 74.75 \\
DSConv & \ourfusion & 64.72 & 71.65 & 54.68 & 63.24 & 70.48 & 72.95 & 66.29 \\
\midrule
\ourmodel & \ourfusion & \textbf{73.90} & \textbf{79.94} & \textbf{67.80} & \textbf{77.65} & \textbf{82.49} & \textbf{83.86} & \textbf{77.61} \\
\bottomrule
\end{tabular}
}
\label{tab:abalation}
\end{table}

\begin{figure}[t]
	\centering
\includegraphics[width=1\linewidth]{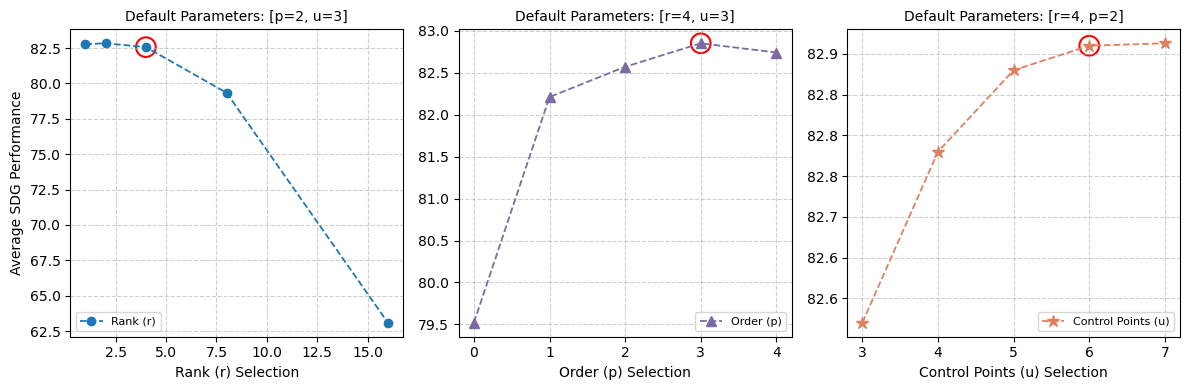}
	\caption{
\textbf{Ablation Study on Parameter Selection}. Red circles distinctly highlight and emphasize the chosen parameters for clear visualization.
    }
    \vspace{-1em}
	\label{fig:abaltion}
\end{figure}

\smallskip
\noindent
\textbf{Ablation Study of Component Parameters.}  
\figureautorefname~\ref{fig:abaltion} presents an ablation analysis of the key parameters using a grid search approach. Specifically, we examine the rank $r $ of \ourmodel~and the order $p $ and number of control points $u $ in \ourfusion. The tested values include $r=[1, 2, 4, 8, 16] $, $p=[0, 1, 2, 3, 4] $, and $u=[3, 4, 5, 6, 7] $. We report the results under the default settings $r=4, p=2, u=3$. We conduct this abaltion study under the DG setting using DECO~\cite{xia2024generalizing}.

Notably, the selection of $r$ adheres to the \textit{elbow rule}, where $r=4$ represents an inflection point that balances parameter reduction with maintaining high performance. For $p=3$, the selection is directly guided by the configuration that yields the best observed performance. For $u $, the results indicate a plateau effect beyond $u=6$, suggesting that this value sufficiently captures the control complexity without leading to overfitting or unnecessary computational overhead. In summary, the best parameters from the grid search is $r=4, p=3, u=6$.

\subsection{ Visualization of Learned Priors }
\label{sec:visualize}

\begin{figure}[t]
	\centering
\includegraphics[width=1\linewidth]{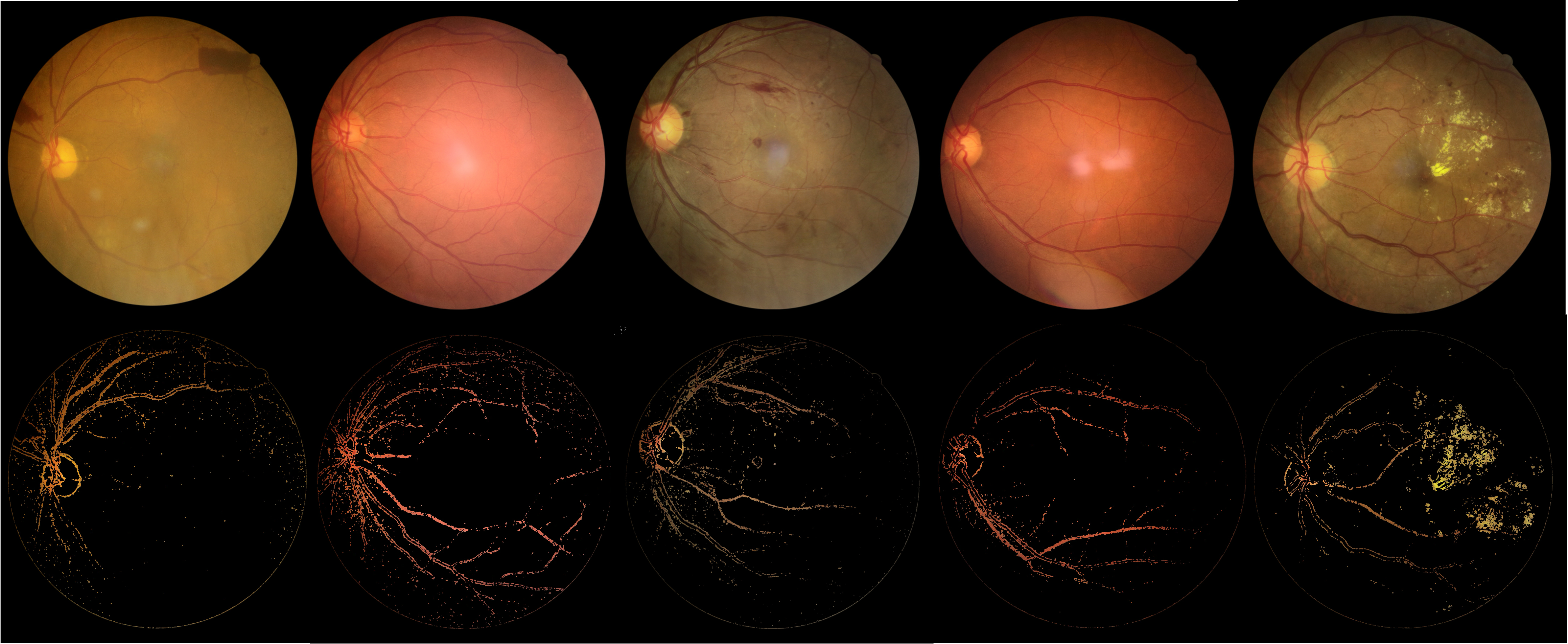}
	\caption{
   \textbf{Visualization of Learned Structural Priors.} The first row presents the retinal imaging inputs, while the second row showcases the corresponding feature visualization results, illustrating the extracted structural priors.
    }
    \vspace{-1em}
	\label{fig:structure}
\end{figure}

Our proposed \ourmodel~is designed to learn structural priors that enhance model generalization. This raises an intriguing question: \textit{what do these learned structural priors look like visually?} To explore this, we visualize the underlying feature map of \ourmodel~after \ourfusion~but before the final fusion step. The feature map is processed by computing column-wise averages and applying a Gaussian filter to suppress activation noise. Subsequently, the \textit{Reds} colormap is applied to the processed feature map to facilitate the clear interpretation of the structure of the vessel. As shown in  \figureautorefname~\ref{fig:structure}, \ourmodel~can capture useful structural features, such as the faint contours of vessel structures, the outline of the eyeball, and discernible lesions (\textit{e.g.,} in the rightmost image). These insights highlight the capacity to capture meaningful structural features for DR grading. However, the visualization also reveals certain limitations, including the omission of fine vessel details and incomplete structural representations, which highlights the need to refine the learned priors to better capture and recognize complex structures in future studies.


\section{Conclusion}

In this work, we present \textit{Low-rank Adaptive Structural Priors (LoASP)}, an innovative plug-and-play module designed for domain generalization (DG), aimed at tracking the critical challenge of domain shifts in diabetic retinopathy (DR) diagnosis. Leveraging the invariant diagnostic features critical for DR classification, including vessel and lesion structures, \ourmodel~effectively learns structural priors and adaptively integrates them with existing approaches, thereby enhancing generalization to unseen domains while preserving computational efficiency and interpretability.  Extensive experiments on diverse datasets validate its effectiveness, showing seamless integration with existing DG methods and improvements in their performance. We anticipate that this research will pave the way for exciting new avenues of exploration, inspiring future innovations in domain generalization for diabetic retinopathy diagnosis and beyond.

\section*{Acknowledgement}
This work is supported by the National Natural Science Foundation
of China (No. 62366005)



 \small \bibliographystyle{ieeetr} \bibliography{main}

\end{document}